\documentclass[12pt,reqno]{amsart}
\pdfoutput=1 
\usepackage{amsmath,bbm}
\usepackage{latexsym}
\usepackage{amsfonts}
\usepackage{amssymb}
\usepackage{color}
\usepackage{graphicx}
\usepackage{url}
\usepackage{enumerate}
\usepackage{tikz}
\usetikzlibrary{shadings, intersections, calc, plotmarks}
\usepackage{marginnote}
\usepackage{stackrel}
\usepackage{empheq}

\usepackage{geometry}
\xdefinecolor{tumblue}     {RGB}{0,101,189}
\xdefinecolor{tumgreen}    {RGB}{162,173,  0}
\xdefinecolor{tumred}      {RGB}{229, 52, 24}
\xdefinecolor{tumivory}    {RGB}{218,215,203}
\xdefinecolor{tumorange}   {RGB}{227,114, 34}
\xdefinecolor{tumlightblue}{RGB}{152,198,234}

\newtheorem{proposition}{Proposition}
\newtheorem{theorem}{Theorem}
\newtheorem*{theorem*}{Theorem}
\newtheorem{lemma}{Lemma}
\newtheorem{corollary}{Corollary}
\newtheorem*{corollary*}{Corollary}

\newtheorem{definition}{Definition}

\newcommand{\id}{{\rm{id}}} 

\newcommand{\cH}{\mathcal{H}}
\newcommand{\R}{\mathbbm{R}}

\newcommand{\C}{\mathbbm{C}}
\newcommand{\N}{\mathbbm{N}}

\newcommand{\F}{\mathbbm{F}}

\newcommand{\cM}{\mathcal{M}}
\newcommand{\1}{\mathbbm{1}}

\def\>{{\rangle}}
\def\<{{\langle}}

\newcommand{\be}{\begin{equation}}
\newcommand{\ee}{\end{equation}}
\newcommand{\bea}{\begin{eqnarray}}
\newcommand{\eea}{\end{eqnarray}}

\newcommand{\tr}[1]{\mathrm{tr}\left[#1\right]} 

\newcommand{\Tr}{\mathrm{tr}}
\newcommand{\rank}[1]{\mathrm{rank}\left[#1\right]}
\newcommand{\norm}[1]{\left\lVert #1 \right\rVert}
\newcommand{\unorm}[1]{\left\vert\hspace*{-1.2pt}\left\vert\hspace*{-1.2pt}\left\vert #1 \right\vert\hspace*{-1.2pt}\right\vert\hspace*{-1.2pt}\right\vert}

\begin{document}

\title[Partial trace relations beyond normal matrices]{Partial trace relations\\ beyond normal matrices}

\author[Rico]{Pablo Costa Rico$^{1,2}$}
\email{pablo.costa@tum.de}
\author[Wolf]{Michael M. Wolf$^{1,2,3}$}
\email{m.wolf@tum.de}
\address{$^1$ Department of Mathematics, Technical University of Munich}
\address{$^2$ Munich Center for Quantum
Science and Technology (MCQST),  M\"unchen, Germany}
\address{$^3$ Pembroke College,  Cambridge, United Kingdom}

\begin{abstract} We investigate the relationship between partial traces and their dilations for general complex matrices, focusing on two main aspects: the existence of (joint) dilations and norm inequalities relating partial traces and their dilations. Throughout our analysis, we pay particular attention to rank constraints. We find that every pair of matrices of equal size and trace admits dilations of any rank larger than one. We generalize Audenaert's subadditivity inequality to encompass general matrices, multiple tensor factors, and different norms. A central ingredient for this is a novel majorization relation for Kronecker sums. As an application, we extend the interval of Werner states in which they are provably 2-undistillable in any dimension $d\geq4$. We also prove new Schmidt-number witnesses and $k$-positive maps.
\end{abstract}
\maketitle
\tableofcontents
\newpage

\section{Introduction}\label{sec:intro}

Marginalization is a fundamental concept that is used wherever product spaces appear---especially in multivariate probability theory and the quantum theory of composite systems. In the latter field, the operation of marginalization as well as its result is called the \emph{partial trace}. Occasionally, this can also be found in other fields \cite{otherptsources} like  
statistical meta-analysis \cite{metaanalysis} or optimal design theory \cite{optdesign}.

In this work, we investigate the partial trace from the perspective of matrix analysis, aiming at a better understanding of the relations between a matrix and its partial traces. Rather than restricting our attention to positive or normal matrices, we deliberately consider general matrices, with the rank as the only occasional constraint. While for positive matrices many results are known and meanwhile basic textbook material, the case of general matrices is significantly less explored. Our motivation for this broader focus stems not only from scientific curiosity but also from the observation that in certain applications, a deeper understanding of partial trace relations beyond the class of normal matrices may be beneficial---and in some cases, even essential. This concerns in particular the distillability problem for Werner states (and with it the notorious NPPT-bound entanglement question \cite{5openproblems,NPPTBE1,NPPTBE2}), which has an equivalent formulation in terms of partial trace relations for rank-constrained but otherwise general matrices. Several other problems lie in close vicinity or are even mathematically equivalent: tensor-stable positive maps \cite{tensorstablepos}, (super-)activation of the distillable entanglement \cite{ActivatingDist} and of the quantum channel capacity with two-way classical communication, or $k$-positivity of certain positive but not completely positive maps (cf. \cite{chen2025symmetryreductiontestingkblockpositivity}).

\subsection*{Outline} To help the reader decide which parts to read, we provide a brief outline of the basic structure and main results of the paper: 

Sec.\ref{sec:ptrange} sets the notation and introduces the partial trace together with the two main properties it preserves: the trace and the sector of the numerical range. 

In Sec.\ref{sec:dilations} we reverse the perspective and discuss \emph{dilations} of matrices to larger ones from which they can be obtained as partial traces. Prop.\ref{prop:structdil} shows that there is no nontrivial constraint for the existence of dilations that are normal, unitary, idempotent, or nilpotent. Props.\ref{prop:Cpartials},\ref{prop:purification} provide necessary and sufficient conditions for the existence of a joint rank-one dilation of two matrices in terms of their Jordan structure. Props.\ref{prop:rank2dilation},\ref{prop:higherrdilations} on the other hand show that for any larger rank there is no nontrivial constraint for the existence of a joint dilation. 

In Sec.\ref{sec:norms} we prove several inequalities that relate the norms of a matrix and its partial traces. Thm.\ref{prop:Agen} does this for general unitarily invariant norms and serves as a template that is then made concrete for Ky Fan $k$-norms and Schatten $p$-norms in subsections Sec.\ref{subsec:KyFan} and Sec.\ref{subsec:Schattenp}, respectively. The main technical Lemma that allows handling non-normal matrices is a majorization result for Kronecker sums presented in Lem.\ref{Lem:Kronsum}. Finally, Sec.\ref{subsec:Werner} and Sec.\ref{sec:Schmidt} will discuss some applications:  the $2$-distillability problem for Werner states and the construction of  Schmidt-number witnesses and $k$-positive maps. 

\subsection*{Related work} Under the assumption of positivity, prior work abounds. Arguably, the most important dilation result is Stinespring's dilation theorem \cite{Stinespring} and the most famous relation between a positive operator and its partial traces is the strong subadditivity of the von Neumann entropy \cite{SSA}.  Beyond those `classic results', more recently, numerous operator inequalities \cite{Lin2023,Choi03082018,ando2014matrix,Li22112023} as well as scalar inequalities \cite{9978926,Fu_Lau_Tam_2021} have been shown to relate positive operators and their partial traces. Moreover, various eigenvalue inequalities have been proven in \cite{Bravyi,Higuchi1,Higuchi2,Franz02} and a general recipe for obtaining such inequalities was established by Klyachko in \cite{klyachko2004}.

Closest to our work are two norm inequalities: in \cite{Audenaert} Audenaert proved that 
\begin{equation}\label{eq:Audineq1}
    \norm{\Tr_A[M]}_p+\norm{\Tr_B[M]}_p\leq \norm{M}_p + \norm{M}_1
\end{equation} for all positive matrices $M$ and all Schatten $p$-norms for $p\in[1,\infty]$. Moreover, in \cite{Rico2025} it was shown that
\begin{equation}\label{eq:Ricosineq}
    \norm{\Tr_A[M]}_2^2+\norm{\Tr_B[M]}_2^2\leq r \norm{M}_2^2 + \frac{1}{r}|\tr{M}|^2
\end{equation} holds for all normal matrices $M$ of rank $r$. Obtaining inequalities similar to those in Eqs.(\ref{eq:Audineq1},\ref{eq:Ricosineq}) but without assuming positivity or normality will be the goal of Sec.\ref{sec:norms}.

Whether Eq.(\ref{eq:Ricosineq}) also holds beyond normal matrices is an open problem that is especially relevant for $r=2$ as this case turns out to be equivalent to the $2$-distillability problem for Werner states \cite{Rico2025}. This problem has been studied in numerous papers \cite{SIO2025152,5508622,QIAN2021139,e20080588,5openproblems,PRAQi25,Wang_2022,e18060216} but remains unresolved. So far, proofs only exist under restrictive assumptions and do not extend to general matrices.

\section{Partial traces and numerical ranges}\label{sec:ptrange}
We will write $\cM_d$ and $\cM_{d_A,d_B}$ for the set of complex $d\times d $ and $d_A\times d_B$ matrices, respectively, and denote by $\cM_{d_A\otimes d_B}:=\cM_{d_A}\otimes\cM_{d_B}$ the Kronecker product with respect to which partial traces are defined. It is useful to have two different, but equivalent, definitions of partial traces in mind: on the one hand, we can regard every $M\in\cM_{d_A\otimes d_B}$ as a block matrix $M\equiv[M_{ij}]$ of $d_A\times d_A$ blocks $M_{ij}\in\cM_{d_B}$. The two partial traces $M_A\equiv\Tr_B[M]\in\cM_{d_A}$ and $M_B\equiv\Tr_A[M]\in\cM_{d_B}$ are then defined via $(M_A)_{ij}:=\tr{M_{ij}}$ and $M_B:=\sum_{i=1}^{d_B} M_{ii}$, respectively. On the other hand, we can view the partial traces as elements of algebras of vector space endomorphisms that satisfy \begin{equation}
    \tr{M_A X}=\tr{M(X\otimes\1_{d_B})},\qquad \tr{M_B Y}=\tr{M(\1_{d_A}\otimes Y)},\label{eq:prtdef}
\end{equation} for all $X\in\cM_{d_A}$, $Y\in\cM_{d_B}$. This perspective makes it obvious that there is no conceptual difference between the two partial traces.  By iteration, both definitions have a straightforward generalization to more than two factors.

While the operator algebraic point of view of Eq.(\ref{eq:prtdef}) allows one to think of the partial traces in a way that does not depend on the choice of the local bases, the block matrix perspective is clearly basis dependent. However, all the properties discussed in this paper will be independent of the choice of orthonormal bases.   

The simplest relation between a matrix and its partial traces is clearly that 
\begin{equation}
    \tr{M_A}=\tr{M_B}=\tr{M}.\label{eq:tracepreservation}
\end{equation}
Another property that is preserved under the partial trace, at least in the weak sense of an inclusion, is the convex cone spanned by the numerical range.
Recall that the \emph{numerical range} of a square matrix $M$, when viewed as an endomorphism on an inner product space, is given by $$W(M):=\{\langle \psi,M\psi\rangle\;|\;\|\psi\|=1\}.$$ 
\begin{proposition}[Partial trace and numerical range]
    If $M\in\cM_{d_A\otimes d_B}$ has partial traces $\rm{tr}_B[M]=A$ and $\rm{tr}_A[M]=B$, then 
    \begin{equation}\label{eq:numrangeincl}
        W(A)\subseteq d_B\;W(M)\qquad\text{and}\qquad W(B)\subseteq d_A\;W(M).
    \end{equation}
\end{proposition}
\begin{proof} We show the first inclusion -- the second one is analogous.
    Let $\varphi\in\C^{d_A}$ be any unit vector, and $\psi_i:=\varphi\otimes e_i$ where $(e_i)_{i=1}^{d_B}$ is an orthonormal basis of $\C^{d_B}$. Then
    \begin{eqnarray}
\langle\varphi,A\varphi\rangle&=&d_B\sum_{i=1}^{d_B}\frac{1}{d_B}\langle\psi_i,M\psi_i\rangle\nonumber\\
&\in& d_B\; {\rm{conv}}\big(W(M)\big)= d_B W(M),\nonumber
    \end{eqnarray}
    where the final equality uses the Toeplitz-Hausdorff theorem by which the numerical range of a matrix is always convex and compact.
\end{proof}
Note that Eq.(\ref{eq:numrangeincl}) generalizes two well-known properties of the partial trace, namely that it preserves hermiticity and positivity. This is implied by Eq.(\ref{eq:numrangeincl}) since hermiticity and positivity are equivalent to the numerical range being a subset of $\R$ and $\R_+$, respectively. 

\section{Dilations}\label{sec:dilations}
We call a matrix $M$ a \emph{dilation} of a given matrix $A$ if $A$ is a partial trace of $M$. In this section, we analyze the existence of dilations with specific properties.

\subsection{Structured dilations}

While hermiticity and positivity are preserved under the partial trace, other properties like normality, unitarity,  nilpotency or idempotency are not conserved to any degree. In fact, every matrix can be obtained as the partial trace of a matrix that is normal or even unitary and, under simple conditions, also from  nilpotent or idempotent matrices:
\begin{proposition}[Structured dilations]\label{prop:structdil} For every $A\in\cM_{d}$ it holds that: 
\begin{enumerate}[(i)]
    \item There is a normal matrix $M\in\cM_{d\otimes 2}$ with $\rm{tr}_B[M]=A$.
    \item There is a unitary matrix $M\in\cM_{d\otimes m}$ with $\rm{tr}_B[M]=A$ if we choose $m$ to be any even number for which $m\geq\|A\|_\infty$.
    \item There is a nilpotent matrix $M\in\cM_{d\otimes 2}$ with $\rm{tr}_B[M]=A$ iff  $\tr{A}=0$.
    \item There is an idempotent matrix $M\in\cM_{d\otimes m}$ with $\rm{tr}_B[M]=A$ for some $m\in\mathbbm{N}$ iff either $A=0$ or $\tr{A}\in\mathbbm{N}$. In this case, we can choose $m=\tr{A}\rank{A}$.
\end{enumerate}
\end{proposition}
\begin{proof} (i) The matrix $M=\tfrac12(A+A^*)\otimes|1\rangle\langle 1|+\tfrac12(A-A^*)\otimes|2\rangle\langle 2|$ has the right partial trace and is normal since it is block-diagonal with two normal blocks: one hermitian and one anti-hermitian.\footnote{Here an in the following, the 'ket´ $|j\rangle$ denotes the $j$'th element of an orthonormal basis.}

(ii) Consider the singular value decomposition $A=U{\rm diag}(s_1,\ldots,s_{d})V$. With $\varphi_k:=\arccos(s_k/m)$ we can write every singular value as $s_k=\tfrac{m}{2}(e^{i\varphi_k}+e^{-i\varphi_k})$. Using the unitaries $W_\pm:=U{\rm diag}(e^{\pm i\varphi_1},\ldots,e^{\pm i\varphi_d})V$ we can define $$M:=\sum_{j=1}^{m/2} W_+\otimes|2j-1\rangle\langle 2j-1|+W_-\otimes|2j\rangle\langle 2j|.$$ By construction, $M$ has the right partial trace and it is unitary as it is block-diagonal with unitary blocks.

(iii) Since every nilpotent matrix $M$ is traceless, the condition $\tr{A}=\tr{{\rm tr}_B[M]}=\tr{M}=0$ is necessary, irrespective of the dimension of the space that is traced out. In order to see that it is also sufficient, we use that  
by Cor.2 in \cite{Fillmore_diagonal}, any matrix is unitarily equivalent to a matrix with constant main diagonal. That is, up to a unitary change of basis, which does not alter any of the considered properties, we have that $A=L+U$ where $L$ and $U$ are matrices that are strictly lower and upper triangular, respectively. Hence, we can choose $M:=L\otimes|1\rangle\langle1|+U\otimes|2\rangle\langle2|$, block-diagonal with nilpotent blocks.

(iv) Since an idempotent matrix $M$ can only have eigenvalues $0$ and $1$, its trace has to be a non-negative integer and it is zero iff $M=0$. So $\tr{A}=\tr{M}\in\mathbbm{N}$ is necessary for any non-zero $A$. To prove sufficiency,
 we will use a composite ancilla system $\C^{d_B}=\C^{d_{B_1}}\otimes\C^{d_{B_2}}$ with $d_{B_1}=\rank{A}$ and $d_{B_2}=\tr{A}$. Following Prop.\ref{prop:purification}(ii) there are vectors $\varphi,\psi\in\C^d\otimes \C^{d_{B_1}}$ s.t. $\langle\psi,\varphi\rangle=1$ and $$A={\rm tr}_{B_1}\big[\tr{A}|\varphi\rangle\langle\psi|\big]={\rm tr}_{B}\big[\underbrace{|\varphi\rangle\langle\psi|\otimes\1_{d_{B_2}}}_{=:M}\big].$$
\end{proof}

\subsection{Joint dilations of rank $1$}

In this subsection, we analyze partial traces of rank-$1$ matrices and (joint) rank-$1$ dilations of matrices.
To this end, we need the following concept:

\begin{definition}[Flanders-similarity]\label{def:Flanders}
    We call two square matrices $A$ and $B$, which are not necessarily of the same size, `\emph{Flanders-similar}' if they have identical Jordan block structure associated with all non-zero eigenvalues, and in addition satisfy the following: If $a_1\geq a_2\geq\ldots$ are the sizes of the Jordan blocks of $A$ that correspond to eigenvalue $0$ and $b_1\geq b_2\geq\ldots$ are the sizes of the respective Jordan blocks of $B$ and the shorter sequence is padded with zeros, then $|a_i-b_i|\leq 1$ for all $i$.\footnote{The sequences $a$ and $b$ are called the \emph{Segré characteristics} at $0$ of $A$ and $B$, respectively.}
\end{definition}
So in particular, similar matrices are Flanders-similar, and for invertible matrices, the two concepts are identical. In general, however, Flanders-similar matrices do not necessarily have equal ranks. Furthermore, Flanders-similarity is not an equivalence relation as it lacks transitivity.
\begin{proposition}[Partial traces of rank-one matrices]\label{prop:Cpartials}
    If $M\in\cM_{d_A\otimes d_B}$ has rank one, then its partial traces $M_A:=\rm{tr}_B[M]$ and $M_B:=\rm{tr}_A[M]$ are Flanders-similar.
\end{proposition}
\begin{proof}
    W.l.o.g. we assume that $d_B\leq d_A$ and write $M=|\Psi_1\rangle\langle\Psi_2|$. Defining $|\Omega\rangle:=\sum_{i=1}^{d_B}|i\rangle\otimes|i\rangle$, we can set $|\Psi_i\rangle=(X_i\otimes\1)|\Omega\rangle$ for suitable $X_1, X_2\in\cM_{d_A,d_B}$. Then
    \begin{equation}\label{eq:partialXCAB}
        {\rm tr}_A[M]=\big(X_2^* X_1\big)^T \quad\text{and}\quad {\rm tr}_B[M]=\big(X_1 X_2^*\big).
    \end{equation}
    The result then follows from the fact that transposition leads to a similar matrix \cite{Taussky1959OnTS} so that $\big(X_2^* X_1\big)^T$ has the same Jordan blocks as $\big(X_2^* X_1\big)$ and interchanging the order in a matrix product leads to a Flanders-similar matrix \cite{Flanders}, which finally relates $\big(X_2^* X_1\big)$ to $\big(X_1 X_2^*\big)$.
\end{proof}
Prop.\ref{prop:Cpartials} can be reversed in the following way:
\begin{proposition}[Purification]\label{prop:purification}\ 
\begin{enumerate}[(i)]
    \item For any two Flanders-similar matrices $A\in\cM_{d_A}$ and $B\in\cM_{d_B}$ there exists a rank-one matrix $M\in\cM_{d_A\otimes d_B}$ s.t. $\rm{tr}_A[M]=B$ and $\rm{tr}_B[M]=A$.
    \item If $A\in\cM_{d_A}$ has rank $r$ then there is a rank-one matrix $M\in\cM_{d_A\otimes d_B}$ such that $A={\rm tr}_B[M]$ iff $d_B\geq r$.
\end{enumerate}    
\end{proposition}
\begin{proof}
   \begin{enumerate}[(i)]
   \item Following \cite{Flanders}, for any two Flanders-similar matrices $A,B$ there are matrices $Y_1, Y_2\in\cM_{d_A,d_B}$ s.t. $A=Y_1 Y_2^*$ and $B=Y_2^* Y_1$. Moreover, there is an invertible $S\in\cM_{d_B}$ for which $SBS^{-1}=B^T$ \cite{Taussky1959OnTS}. Setting $X_1:= Y_1 S^{-1}$, $X_2^*:=S Y_2^*$ the construction preceding Eq.(\ref{eq:partialXCAB}) yields the desired $M$.
   \item From Prop.\ref{prop:Cpartials} and (i) we know that a matrix $M$ as described exists iff there is a $B\in\cM_{d_B}$ that is Flanders-similar to $A$. So the questions is: in which dimensions are there Flanders-similar matrices? By Def.\ref{def:Flanders} the smallest dimension is $d_A$ minus the number of Jordan blocks of $A$ associated to eigenvalue zero. This, however, is nothing but the rank $r$ of $A$ as each of those Jordan blocks reduces the rank by exactly $1$. Every dimension larger than $r$ is possible as well since we can always attach any number of one-dimensional Jordan blocks associated to eigenvalue zero.
   \end{enumerate}
\end{proof}

\subsection{Joint dilations of rank $r\geq 2$}
The results of the previous subsection impose severe restrictions on a pair of matrices if a joint dilation of rank one is to exist. In the following, we will see that these restrictions disappear almost entirely for any rank larger than one.
\begin{proposition}[Simultaneous rank-two dilation]\label{prop:rank2dilation}
    For any two matrices $A,B\in\cM_{d}$  there exists a rank-two matrix $M\in\cM_{d\otimes d}$ s.t. $\rm{tr}_A[M]=B$ and $\rm{tr}_B[M]=A$ iff $\tr{A}=\tr{B}$.
\end{proposition}
\begin{proof} As partial traces preserve the trace, $\tr{A}=\tr{B}$ is clearly necessary. So we will assume equal traces and now aim at proving sufficiency.

    Reusing the argument leading to Eq.(\ref{eq:partialXCAB}) we can reformulate the proposition such that it is equivalent to the existence of a solution to 
    \begin{equation}\label{eq:rank2XY}
         B^T= X_2^*X_1+Y_2^*Y_1,\qquad A=X_1 X_2^*+Y_1 Y_2^*,
    \end{equation}
    with indeterminate matrices $X_i,Y_j\in\cM_{d}$. As the difference $B^T-A$ is traceless it can be expressed as a commutator $B^T-A=KL-LK$ of two matrices $K,L\in\cM_d$ according to Shoda's theorem \cite{Shoda}. The choice $X_2^*:=K, X_1:=L,Y_2^*=\1, Y_1=A-LK$ then solves Eq.(\ref{eq:rank2XY}).
       \end{proof}
It is worth noting that Prop.\ref{prop:rank2dilation} does not hold under positivity constraints, i.e. if $A,B\geq 0$ then there need not exist a simultaneous dilation $M\geq 0$ that has rank lower than the local Hilbert space dimension, as exemplified in the following: \vspace*{5pt}

{\emph{Example}:} Consider $A,B\in\cM_d$ where $A=|0\rangle\langle0|$ and $B=\1/d$. If $M\in\cM_{d\otimes d}$ is positive semidefinite of rank $r$ with partial traces $A$ and $B$, then there are matrices $(X_i\in\cM_d)_{i=1}^r$ s.t.
$A=\sum_i X_i X_i^*$ and $B=\sum_i X_i^* X_i$. The choice of $A$ implies that $X_i=|0\rangle\langle x_i|$. Consequently, $B=\sum_{i=1}^r |x_i\rangle\langle x_i|$, which requires $r\geq d$. In fact, $r=d$ is necessary and sufficient as $M=A\otimes B$ is the only positive dilation for this example.\vspace*{5pt}

If there are significant differences between the dimensions $d_A$ and $d_B$ as well as between the ranks, then a joint low-rank dilation cannot exist:

\begin{proposition}[Dimension constraints for joint dilations] If $A\in\cM_{d_A}$ and $B\in\cM_{d_B}$ are the partial traces of $M\in\cM_{d_A\otimes d_B}$, then \begin{equation}
    \max{\{{\rm rank}(A),{\rm rank}(B)}\}\leq {\rm rank}(M)\min\{d_A,d_B\}.
\end{equation}\label{prop:jointdildimconst}    
\end{proposition}
\begin{proof}
    Using again the argument that led to Eq.(\ref{eq:partialXCAB}), we know that a joint dilation of rank $r$ implies the existence of matrices $X_i,Y_i\in\cM_{d_A,d_B}$ s.t.
    $$B^T=\sum_{i=1}^r Y_i^* X_i,\qquad A=\sum_{i=1}^r X_i Y_i^*.$$ Since ${\rm rank}(Y_i^* X_i),{\rm rank}( X_i Y_i^*)\leq \min\{d_A,d_B\}$ each of the partial traces can have rank at most $r$ times the minimal dimension.
\end{proof}
{\emph{Example}:} If $A$ has full rank, then any dilation $M$ has ${\rm rank}(M)\geq d_A/d_B$.\vspace*{5pt}  

However, if we seek dilations with large rank, then there is no constraint even if the dimensions differ: 
\begin{proposition}[Large rank dilations]\label{prop:higherrdilations}
    If $A\in\cM_{d_A}$ and $B\in\cM_{d_B}$ have a joint dilation $M\in\cM_{d_A\otimes d_B}$, then they have a joint dilation of any rank $r$ with ${\rm rank}(M)\leq r\leq d_A d_B$. 
\end{proposition}
\emph{Remark:} In particular, if $d_A=d_B=d$ then there are joint dilations of any rank $2\leq r\leq d^2$.
\begin{proof}
    To construct a dilation of maximal rank, we denote by $S_d\in\cM_d$ a matrix that acts as a cyclic shift on an orthonormal basis of $\C^d$. This satisfies ${\rm rank}(S_d)=d$ and $\tr{S_d}=0$. Consequently, $M_t:=M+t S_{d_A}\otimes S_{d_B}$ has full rank for almost every $t\in\R$ (by the lower semicontinuity of the rank). Moreover, $\Tr_A{[M_t]}=\Tr_A[M]+t\;\tr{S_{d_A}}S_{d_B}=A$ and similarly $\Tr_B{[M_t]}=B$ for every $t$. To obtain dilations of smaller rank, we can follow the same argument where we replace $S_{d_A}\otimes S_{d_B}$  successively by the matrices obtained from it by setting more and more of its rows to zero. 
\end{proof}

\section{Norm inequalities}\label{sec:norms}
We will now turn to norm inequalities between a matrix and its partial traces. We consider unitarily invariant norms, which we assume to be defined for matrices of all sizes, and we will denote them in general by $\unorm{\cdot}$ with dual norm $\unorm{\cdot}^*$. Two types of unitarily invariant norms that will be discussed in greater detail are the Schatten $p$-norm $\norm{\cdot}_p$ (i.e., the $l_p$-norm of the singular values for $p\in[1,\infty]$) and the Ky Fan $k$-norm $\norm{\cdot}_{(k)}$ (i.e., the sum of the $k$ largest singular values).

We begin with a slight extension of results from \cite{Rastegin2012, Rico2025}:
\begin{proposition}[Norm inequalities for individual partial traces]\label{prop:singlenorms}
    If $M\in \cM_{d_A\otimes d_B}$ has rank $r$ and partial traces $A={\rm tr}_B[M]$ and $B={\rm tr}_A[M]$, then for every unitarily invariant norm:
    \begin{equation}
        \unorm{A}\leq\nu\unorm{M},\quad \unorm{B}\leq\nu\unorm{M},\quad\text{with } \nu:=\unorm{E}\unorm{\1_r}^*,
    \end{equation}
    where $\1_r$ is the $r\times r$ identity matrix and $E:={\rm diag}(1,0,\ldots,0)$. Specifically,
    \begin{equation}
    \nu=\begin{cases}
r^{1-1/p} & \text{if}\quad  \unorm{\cdot}=\norm{\cdot}_p \\
\max \big\{1,\tfrac{r}{k}\big\} & \text{if}\quad \unorm{\cdot}=\norm{\cdot}_{(k)}
\end{cases}
    \end{equation}
\end{proposition}
\begin{proof}
    We will prove the inequality for $A$---the one for $B$ is proven analogously. Since the trace norm is the largest unitarily invariant norm for which $E$ has norm one (\cite{Bhatia}, p.93), we can bound $\unorm{A}\leq\unorm{E}\norm{A}_1$. Moreover, 
    \begin{eqnarray}
        \norm{A}_1 &=& \sup\big\{|\tr{AX}|\;\big|\;\norm{X}_\infty\leq 1\big\}\nonumber \\ &=& \sup\big\{|\tr{M(X\otimes\1_{d_B})}|\;\big|\;\norm{X}_\infty\leq 1\big\}\nonumber \\ 
        &\leq& \sup\big\{|\tr{MY}|\;\big|\;\norm{Y}_\infty\leq1\big\}=\norm{M}_1 .\nonumber
    \end{eqnarray}
    Finally, we can use duality of the norms (or equivalently, a general Cauchy-Schwarz inequality (\cite{Bhatia}, p.96) ) together with the unitary invariance of the dual norm to arrive at 
    \begin{equation}
        \norm{M}_1=\tr{U\1_r V M}\leq\unorm{U\1_r V}^*\unorm{M}=\unorm{\1_r}^*\unorm{M},
    \end{equation}
    where $U,V$ are unitaries whose existence is guaranteed by the singular value decomposition of $M$ and $\1_r$ is the (embedded) identity matrix.

    For the specified norms, the scalar $\nu$ is derived by using that $E$ has norm $1$ in all considered cases and that $\norm{\cdot}_p^*=\norm{\cdot}_q$ with $1/p+1/q=1$ and $\norm{\cdot}_{(k)}^*=\max\{\norm{\cdot}_\infty,\frac{1}{k}\norm{\cdot}_1\}$ (\cite{Bhatia}, p.90).
\end{proof}
\ \vspace*{5pt}

We proceed by deriving norm inequalities that involve more than a single partial trace and relate their norms to the norm of the dilation. The following proposition will be formulated for an arbitrary unitarily invariant norm and serve as a template for more specific norm inequalities discussed in subsequent subsections. We will denote the symmetric gauge functions that correspond to the norm and its dual by $\phi$ and $\phi^*$, respectively. Moreover, we use the shorthand $(x)_+:=\max\{x,0\}$ for the positive part of a real number.
\begin{theorem}[Norm inequality template]\label{prop:Agen}
    Let $M\in\cM_{d_1\otimes\ldots\otimes d_n}$ have rank $r$ and partial traces $(M_i\in\cM_{d_i})_{i=1}^n$. For any $c\geq 0$ the following holds:
    \begin{eqnarray}
        \sum_{i=1}^n\unorm{M_i} &\leq& c\;\|M\|_1+\kappa(c)\unorm{M}\;,\\
        \kappa(c) &:=& \sup\phi^*\big(\Lambda_1,\ldots,\Lambda_r\big),
    \end{eqnarray}
    where $\Lambda$ is a vector with $\prod_{i=1}^n d_i$ decreasingly ordered components of the form $\big(-c+\sum_{i=1}^n \lambda_{j_i}^{(i)}\big)_+$ and the supremum is taken over all $n$-tuples of vectors $\big(\lambda^{(i)}\in\R_+^{d_i}\big)_{i=1}^n$ satisfying $\phi^*(\lambda^{(i)})\leq 1$.\footnote{That is, every single component of $\Lambda$ corresponds to exactly one multiindex $(j_1,\ldots,j_n)$, where $j_i\in\{1,\ldots,d_i\}$. The value of this component is then $\big(-c+\sum_{i=1}^n \lambda_{j_i}^{(i)}\big)_+$. Out of the $\prod_{i=1}^n d_i$ components of $\Lambda$, only the $r$ largest ones are used.}
\end{theorem}
 
\begin{proof}
By definition of the dual norm, we can write $\unorm{M_i}=\sup|\tr{M_i C_i}|$ taken over all $C_i$ with $\unorm{C_i}^*\leq 1$. Defining, up to rearrangement, $C_{(i)}:=C_i\otimes\1$ so that $\tr{M_i C_i}=\tr{M C_{(i)}}$ and introducing   an orthonormal projection $P$ of rank $r$ so that $MP=M$, we get for an arbitrary $Z$ by virtue of the triangle inequality that
\begin{eqnarray}
    \sum_{i=1}^n\unorm{M_i} &\leq& \sup \left|\tr{MP\left(Z+\sum_{i=1}^n C_{(i)}\right)}\right|+|\tr{MZ}|\\
    &\leq & \unorm{M} \sup \unorm{P\left(Z+\sum_{i=1}^n C_{(i)}\right)}^* +\|M\|_1\|Z\|_\infty,\label{eq:kappasupc}
\end{eqnarray}
where duality of the norms was used for both terms in the second inequality. The dual norm can be bounded by using that for any matrix $X$ the following weak submajorization holds for the singular values (cf. Thm.IV.2.5 in \cite{Bhatia})
$$\big(\sigma_j(PX)\big)\prec_{\mathrm{w}}\big(\sigma_j(P)\sigma_j(X)\big)=\big(\sigma_1(X),\ldots,\sigma_r(X)\big),
$$ which implies $\unorm{PX}^*\leq \phi^*\big(\sigma_1(X),\ldots,\sigma_r(X)\big)$.

The matrix $Z$ will now be chosen such that it diminishes each singular value of the sum $\sum_i C_{(i)}$ by the largest value not exceeding $c$ while not changing the unitaries of the singular value decomposition. With that we get $\norm{Z}_\infty\leq c$ and the singular values of the matrix $X=Z+\sum_i C_{(i)}$ become $$\sigma_j(X)=\left(-c+\sigma_j\bigg(\sum_{i=1}^n C_{(i)}\bigg)\right)_+ = g\left(\sigma_j\bigg(\sum_{i=1}^n C_{(i)}\bigg)\right),$$ where $g:\R\rightarrow\R,\; g(x):=(x-c)_+$. As $\phi^*$, like every symmetric gauge function, is Schur convex and increasing and $g$ is convex and increasing, the composition $\phi^*\circ g$, where $g$ is applied componentwise, is an increasing Schur-convex map (cf. \cite{MOmajo}, p.91). Using Lemma \ref{Lem:Kronsum} this implies that $$\phi^*\big(\sigma_1(X),\ldots,\sigma_r(X)\big)\leq\phi^*\circ g\bigg(\sigma_1\Big(\sum_{i=1}^n |C_{(i)}|\Big),\ldots,\sigma_r\Big(\sum_{i=1}^n |C_{(i)}|\Big)\bigg).$$ 
Finally, we denote the vector of eigenvalues of $|C_i|$ by $\lambda^{(i)}\in\R_+^{d_i}$ and notice that $\unorm{C_i}^* = \phi^*(\lambda^{(i)})$. The proof is then completed by realizing that the $g\Big(\sigma_j\Big(\sum_{i=1}^n |C_{(i)}|\Big)\Big)$ coincide with the $\Lambda_j$ when sorted in decreasing order. 
\end{proof}
It remains to prove the following central ingredient:
\begin{lemma}[Kronecker-sum majorization]\label{Lem:Kronsum}
    For an $n$-tuple of matrices $(C_i\in\cM_{d_i})_{i=1}^n$ define their embeddings into the tensor product space by $C_{(i)}\in\cM_{d_1\otimes\ldots\otimes d_n}$ s.t. $C_{(i)}$ is a tensor product with $C_i$ on the $i'th$ tensor factor and identities elsewhere. The vectors of singular values satisfy the weak submajorization relation
    \begin{equation}
        \bigg(\sigma_j\Big(\sum_{i=1}^n C_{(i)}\Big)\bigg)\prec_{\mathrm{w}} \bigg(\sigma_j\Big(\sum_{i=1}^n |C_{(i)}|\Big)\bigg).
    \end{equation}
\end{lemma}
\emph{Remark}: $\sum_i C_{(i)}$ is called the \emph{Kronecker sum} of the $C_i$'s.
\begin{proof}
    With the polar decomposition $C_i=U_i|C_i|$ and an orthonormal basis $\{|i\rangle\in\C^n\}$ we can write
    \begin{eqnarray}
        \sum_{i=1}^n C_{(i)} &=& L\; R\qquad \text{with}\\
        L &:=& \sum_{i=1}^n U_{(i)} |C_{(i)}|^{\frac{1}{2}} U_{(i)}^*\otimes \langle i|,\nonumber\\
        R &:=& \sum_{i=1}^n U_{(i)} |C_{(i)}|^{\frac{1}{2}} \otimes |i\rangle .\nonumber
    \end{eqnarray}
    Weak submajorization holds iff the corresponding inequality holds for all unitarily invariant norms. We therefore consider an arbitrary unitarily invariant norm $\unorm{\cdot}$ to which we apply the Cauchy-Schwarz inequality (cf. \cite{Bhatia} IX.5) to obtain
    \begin{eqnarray}
        \unorm{\sum_{i=1}^n C_{(i)}} &=& \unorm{LR}\leq\unorm{L^*L}^{\frac{1}{2}}\unorm{R^*R}^{\frac{1}{2}}\\ &=& \unorm{L L^*}^{\frac{1}{2}}\unorm{R^*R}^{\frac{1}{2}}\nonumber\\
        &=& \unorm{U\Big(\sum_{i=1}^n |C_{(i)}|\Big) U^*}^{\frac{1}{2}}\unorm{\sum_{i=1}^n |C_{(i)}|}^{\frac{1}{2}},\nonumber
    \end{eqnarray} where $U:=\otimes_{i=1}^n U_i$. Exploiting the unitary invariance of the norm then gives the claimed result.
\end{proof}

\subsection{Ky Fan $k$-norms} \label{subsec:KyFan}
In this section we will show how to compute the function $\kappa(c)$ of Thm.\ref{prop:Agen} for the case where $\unorm{\cdot}=\norm{\cdot}_{(k)}$ is a Ky Fan $k$-norm, i.e. the sum of the largest $k$ singular values.
\begin{lemma}[Ky Fan $k$-norm optimizers]
    If $\unorm{\cdot}=\norm{\cdot}_{(k)}$ is the Ky Fan $k$-norm, then the supremum in Eq.(\ref{eq:kappasupc}) is attained if  $\lambda^{(i)}=(1,\ldots,1,0,\ldots,0)$ for every $i$ where $1$ appears $\min\{d_i,k\}$ times.

    In the case of equal dimensions $d_i=d$, this implies  for $k\leq d$ that  \begin{equation}
        \Big(\sum_{i=1}^n \lambda_{j_i}^{(i)}\Big)^\downarrow=(n,\ldots,n,n-1,\ldots,n-1,\ldots,\underbrace{n-l,\ldots,n-l}_{\binom{n}{l}(d-k)^l k^{n-l}\text{ times}},\ldots),\label{eq:lambdasortkyFanopt}
    \end{equation}
    with $l\in\{0,\ldots,n\}$. For $k>d$ the expression is the same as for $k=d$, i.e., every component is equal to $n$.
\end{lemma}
\begin{proof}
    The symmetric gauge function of the dual norm of the Ky Fan $k$-norm is given by $\phi^*(x):=\max\{\|x\|_\infty,\frac{1}{k}\|x\|_1\}$ (\cite{Bhatia} p.90) where the occurring norms are the $l_\infty$ and the $l_1$ norm.

    We want to compute $\kappa(c):=\sup\phi^*\big(\Lambda_1,\ldots,\Lambda_r\big)$, where $\Lambda$ is defined by ordering the multiset $\big\{\big(-c+\sum_{i=1}^n \lambda_{j_i}^{(i)}\big)_+\big\}$ and the supremum is taken over all $n$-tuples of vectors $\big(\lambda^{(i)}\in\R_+^{d_i}\big)_{i=1}^n$ satisfying $\phi^*(\lambda^{(i)})\leq 1$.

Note that $\psi(x):=\phi^*(|x_1|^\downarrow,\ldots,|x_r|^\downarrow)$ is again a symmetric gauge function (\cite{Bhatia}, p.90), and thus symmetric, convex, and component-wise increasing for positive components \cite{Mirsky}. If we fix all vectors $\lambda^{(i)}$ but one, the map $\kappa_i:\lambda^{(i)}\mapsto \psi(\Lambda)$ is therefore convex and symmetric. Following \cite{MOmajo} (p.97) it is then Schur convex. As $\kappa_i$ is also component-wise increasing we have (\cite{MOmajo}, p.87) 
\begin{equation}\label{eq:ouewfo}
    \lambda^{(i)}\prec_{\mathrm{w}} v\quad\Rightarrow\quad \kappa_i\big(\lambda^{(i)}\big)\leq \kappa_i(v).
\end{equation}
Choosing $v:=(1,\ldots,1,0,\ldots,0)$, where $1$ appears $\min\{k,d_i\}$ times, we get that $\phi^*(v)\leq 1$ and $v$ weakly submajorizes any $\lambda^{(i)}\in\R_+^{d_i}$ for which $\phi^*(\lambda^{(i)})\leq 1$.  Eq.(\ref{eq:ouewfo}) then implies that the supremum in $\kappa(c)$ is attained if each $\lambda^{(i)}$ is set equal to the corresponding $v$. Inserting this into the l.h.s. of Eq.(\ref{eq:lambdasortkyFanopt}), the r.h.s. follows from elementary combinatorics.
\end{proof}
Inserting the optimizers into Thm.\ref{prop:Agen} for different dimensions and values of $k,c,r$ leads to a plethora of inequalities, some of which are summarized in the following:
\begin{corollary}[Ky Fan $k$-norm inequalities]
    Let $M\in\cM_{d_1\otimes\ldots\otimes d_n}$ have rank $r$ and partial traces $(M_i\in\cM_{d_i})_{i=1}^n$ of equal dimension $d_i=d$. Then
\begin{subequations}
\begin{empheq}[left={\sum_{i=1}^n\norm{M_i}_{(k)} \leq \empheqlbrace}]{align}
  \norm{M}_1+\tilde{\kappa}\norm{M}_{(k)} & \label{eq:kyfancor1}  \\ \label{eq:kyfancor2}
  \norm{M}_1+k\norm{M}_{(k)}  & \quad \text{for } n=2\\
  c\norm{M}_1+(n-c)_+\norm{M}_{(k)} & \quad \text{for } r\leq k, \text{ and }  c\geq 0 \label{eq:kyfancor3}
\end{empheq}
\end{subequations}
where $\tilde{\kappa}:=\begin{cases}
    n d^{n-1}-\frac{d^n-(d-k)^n}{k} &,\text{ if } k\leq d,\\
    (n-1) k^{n-1}&,\text{ if } k\geq d.
\end{cases}$.
\end{corollary}
\begin{proof}
    Eq.(\ref{eq:kyfancor1}): As this bound has to be valid for arbitrary rank, we take all $d^n$ components of the optimizer in Eq.(\ref{eq:lambdasortkyFanopt}) and insert them into the computation of $\kappa(c)$, where we choose $c=1$. Suppose first that $k\leq d$. Then $\Lambda$ gets $\binom{n}{l}(d-k)^l k^{n-l}$ components with value $n-l-1$ for $l\in\{0,n-1\}$. Consequently,
    \begin{equation}
        \phi^*(\Lambda)=\max\bigg\{n-1,\underbrace{\frac{1}{k}\sum_{l=0}^{n-1}(n-l-1)\binom{n}{l}(d-k)^l k^{n-l}}_{=\;n d^{n-1}-\frac{d^n-(d-k)^n}{k}\;=:\;\tilde{\kappa}}\bigg\}
    \end{equation}
    where $\tilde{\kappa}$ is never smaller than $n-1$, which can be seen by noting that the $l=0$ term of the sum already satisfies this bound. If $k>d$, then $\Lambda$ has the same components as for $k=d$, which proves the claimed value for $\tilde{\kappa}$.

    Eq.(\ref{eq:kyfancor2}) follows from Eq.(\ref{eq:kyfancor1}) by inserting $n=2$.

    Eq.(\ref{eq:kyfancor3}) uses that for $r\leq k^n$  the largest $r$ components of the optimal $\Lambda$ are all equal to $(n-c)_+$ so that 
    \begin{equation}
        \phi^*(\Lambda_1,\ldots,\Lambda_r) =\max\Big\{(n-c)_+,\tfrac{r}{k}(n-c)_+\Big\},
    \end{equation} where the first term achieves the maximum if $r\leq k$.
\end{proof}


\subsection{Schatten $p$-norms}\label{subsec:Schattenp}
In this section, we will discuss the derivation of the function $\kappa(c)$ of Thm.\ref{prop:Agen} for the case where $\unorm{\cdot}=\norm{\cdot}_p$ is a Schatten-$p$ norm, i.e. the $l_p$-norm of the vector of singular values. 
We will explicitly work out the norm bounds for some interesting cases of $n,p,d,r,c$ in the main text and provide a way of simplifying the underlying optimization problem more generally in the Appendix.

We begin with a generalization of Audenaert's inequality to arbitrary matrices and more than two tensor factors:
\begin{proposition}[Extended Audenaert inequality]\label{prop:ExtendedAudenaert}
    If $M\in\cM_{d_1\otimes\ldots\otimes d_n}$ has  partial traces $(M_i\in\cM_{d_i})_{i=1}^n$, then for all $p\in[1,\infty]$: 
    \begin{equation}\label{eq:Audgenprop}
        \sum_{i=1}^n\norm{M_i}_p\leq(n-1)\norm{M}_1+\norm{M}_p.
    \end{equation}
\end{proposition}
\begin{proof}
    Note first that it suffices to prove this for $n=2$ as the general case follows from a simple induction on $n$: assuming the statement for all numbers of tensor factors up to $n$, we can write
    \begin{eqnarray}
        \sum_{i=1}^{n+1}\norm{M_i}_p &=& \Big(\sum_{i=1}^{n}\norm{M_i}_p\Big)+\norm{M_{n+1}}_p\nonumber \\ &\leq& (n-1)\underbrace{\norm{M_{1..n}}_1}_{\leq\norm{M}_1} +\underbrace{\norm{M_{1..n}}_p+\norm{M_{n+1}}_p}_{\leq \norm{M}_1+\norm{M}_p}\nonumber\; ,
    \end{eqnarray} which proves the inequality for $n+1$.

    For $n=2$ the statement follows from Thm.\ref{prop:Agen} together with Audenaert's proof of Lem.1 in\cite{Audenaert}, which we repeat here for completeness: the aim is to compute $\kappa(1)$ for $n=2$, i.e.
       \begin{equation}
         \kappa(1)=\sup\bigg\{\sum_{i=1}^{d_1}\sum_{j=1}^{d_2}(x_i+y_j-1)_+^q\;\Big|\; \sum_{i=1}^{d_1} |x_i|^q=\sum_{j=1}^{d_2}|y_j|^q=1\bigg\}^{1/q},\nonumber \end{equation}
         where $1/p+1/q=1$.
     To this end, Audenaert considers the function $f(x_i):=\big(\sum_{j=1}^{d_2}(x_i+y_j-1)_+^q\big)^{1/q}$. If $\norm{y}_q=1$, this satisfies $f(0)=0$ and $f(1)=1$. Convexity of $f$ then implies that $f(x_i)\leq x_i$ for all $x_i\in[0,1]$. Consequently, on the positive orthant
     \begin{equation}\nonumber
         \sum_{i=1}^{d_1}\sum_{j=1}^{d_2}(x_i+y_j-1)_+^q = \sum_{i=1}^{d_1} f(x_i)^q\leq \sum_{i=1}^{d_1} x_i^q=1.
     \end{equation} As equality is achieved for $x=(1,0,\ldots,0)$ we have $\kappa(1)=1$.
\end{proof}
\begin{corollary}\label{cor:AudenR}
If $M\in\cM_{d_1\otimes\ldots\otimes d_n}$ has rank  $r$ and partial traces $(M_i\in\cM_{d_i})_{i=1}^n$, then for all $p\in[1,\infty]$ and $\gamma\geq 1$: 
    \begin{equation}\label{eq:lowrankprop}
        \sum_{i=1}^n\norm{M_i}_p^\gamma\leq\Big(1+r^{\gamma(1-1/p)}(n-1)\Big)\norm{M}_p^\gamma.
    \end{equation}    
\end{corollary}
\begin{proof} For $\gamma=1$ this follows from Eq.(\ref{eq:Audgenprop}) together with H\"older's inequality: if $P$ is an orthogonal projection onto the range of $M$, then with $1/p+1/q=1$:
$$\norm{M}_1=\norm{PM}_1\leq \norm{P}_q\norm{M}_p=r^{1/q}\norm{M}_p.$$
In order to extend the inequality to $\gamma\geq 1$, we use that according to Prop.\ref{prop:singlenorms} $\norm{M_i}_p\leq r^{1-1/p}\norm{M}_p$. This, together with Eq.(\ref{eq:lowrankprop}) applied with $\gamma=1$, implies 
    $$\Big(\norm{M_i}_p\Big)_{i=1}^n\prec_w\Big(\underbrace{r^{1-1/p}\norm{M}_p,\ldots}_{(n-1) \text{ times}},\norm{M}_p\Big).$$
    The proof is then completed by applying the Schur convex function $f:\R^n_+\ni x\mapsto \sum_{i=1}^n x_i^\gamma$. 
    \end{proof}
As the following example shows, the inequality is tight as long as the rank $r$ does not exceed the largest local dimension:

\emph{Example}: Consider $M=|1\rangle\langle1|^{\otimes(n-1)}\otimes \1_r$. Then for all $i<n$: $M_i=r|1\rangle\langle1|$ have norm $\norm{M_i}_p=r$ and $\norm{M_n}_p=\norm{M}_p=r^{1/p}$. That is, equality holds in Eq.(\ref{eq:lowrankprop}).\vspace*{5pt} 

However, for large rank the inequality in Eq.(\ref{eq:lowrankprop}) can be improved. In particular, if the rank constraint is dropped entirely:
\begin{proposition}
    If $M\in\cM_{d_1\otimes\ldots\otimes d_n}$ has  partial traces $(M_i\in\cM_{d_i})_{i=1}^n$ of equal dimension $d_i=d$, then for all $p\in[1,\infty]$: 
    \begin{equation}\label{eq:largerankprop}
        \sum_{i=1}^n\norm{M_i}_p\leq n\cdot d^{(n-1)(1-1/p)}\norm{M}_p\;.
    \end{equation}
\end{proposition}
\emph{Remark:} The following proof works for unequal dimensions as well -- only the result becomes more cumbersome to state. Equality in Eq.(\ref{eq:largerankprop}) is attained, for example, for $M=\1_d^{\otimes n}$.
\begin{proof}
    For $z^{(i)}\in[0,1]^d, i\in\{1,\ldots,n\}$ define
    \begin{equation}
        F\big(z^{(1)},\ldots,z^{(n)}\big):=\sum_{j_1=1}^d \cdots\sum_{j_n=1}^d\bigg(\sum_{i=1}^n\big(z_{j_i}^{(i)}\big)^{1/q}\bigg)^q.
    \end{equation}
    Fixing all but one argument, we obtain a function $z^{(i)}\mapsto F\big(z^{(1)},\ldots,z^{(n)}\big)=:f(z^{(i)})$ that is concave for any $q>1$ (as it is a sum of concave functions) and symmetric w.r.t. permuting the components of $z^{(i)}$. Therefore, $f$ is Schur-concave (cf.\cite{MOmajo}, p.97) so that the majorization $e:=(1/d,\ldots,1/d)\prec z^{(i)}$ implies $f\big(z^{(i)}\big)\leq f(e)$. Applying this to every single argument of $F$ and utilizing Thm.\ref{prop:Agen} leads to
    \begin{eqnarray}
        \kappa(0) &=&\sup\Big\{F\big(z^{(1)},\ldots,z^{(n)}\big)^{1/q}\;\big|\; z^{(i)}\in[0,1]^d,\;\norm{z^{(i)}}_1=1\Big\}\nonumber\\
        &=& F(e,\ldots,e)^{1/q}\ =\ n\cdot d^{(n-1)/q}.\nonumber
    \end{eqnarray}
\end{proof}

Finally, we will derive a $2$-norm inequality using a different way of reasoning that is only available for rank-one dilations. This inequality appeared first, with a slightly different proof and for $\gamma=2$, in \cite{Rico2025}: 

\begin{proposition}
    If $M\in\cM_{d_A\otimes d_B}$ has rank one and partial traces $A\in\cM_{d_A}$ and $B\in\cM_{d_B}$, then for any  $\gamma\geq 2$:
    \begin{equation}
        \norm{A}_{2}^\gamma + \norm{B}_{2}^\gamma \leq \norm{M}_{2}^\gamma + \big|\tr{M}\big|^\gamma\label{eq:trineqr1}.
    \end{equation}
\end{proposition}
\begin{proof}
    Following the construction in the proof of Prop.\ref{prop:Cpartials}, let $M$ be described by two matrices $X,Y\in\cM_{d_A,d_B}$ so that $A^T=Y^*X$, $B=XY^*$ and $\norm{M}_2=\norm{X}_2\norm{Y}_2$. Denote by $\F$ the \emph{flip} operator that acts on a tensor product of isomorphic spaces as $\F:\psi\otimes \phi\mapsto\phi\otimes\psi$. Using that $(\1-\F)$ is positive semidefinite, we get
    \begin{eqnarray}
        0 &\leq& \tr{(\1-\F)(X\otimes Y)(\1-\F)(X^*\otimes Y^*)}\nonumber\\
        &=&\tr{XX^*\otimes YY^*}+\tr{XY^*\otimes YX^*}-\tr{XY^*YX^*}-\tr{XX^*YY^*}\nonumber\\
        &=& \norm{M}_2^2+|\tr{M}|^2-\norm{A}_2^2-\norm{B}_2^2,
    \end{eqnarray}
    which proves the inequality for $\gamma=2$. In order to generalize it to $\gamma\geq 2$, we use that according to Prop.\ref{prop:singlenorms} $\norm{A}_2,\norm{B}_2\leq\norm{M}_2$. Hence 
    $$\big(\norm{M}_2^2,|\tr{M}|^2\big)\succ_w\big(\norm{A}_2^2,\norm{B}_2^2\big),$$
    so that the proof can be completed by applying the Schur convex function $f:\R^2_+\ni(x_1,x_2)\mapsto x_1^{\gamma/2}+x_2^{\gamma/2}$. 
\end{proof}\vspace*{5pt}

\subsection{Application: $2$-distillability of Werner states}\label{subsec:Werner}
\emph{Werner states} are one-parameter families of density operators $\rho_{\alpha}\in\cM_{d\otimes d}$ of the form \begin{equation}
    \rho_{\alpha}:=\frac{\1+\alpha\F}{d^2+\alpha d},\qquad \alpha\in[-1,1].\label{eq:Werner}
\end{equation}
These are known to be \emph{one-copy undistillable}\footnote{A bipartite quantum state is called $n$-copy undistillable if it is not possible to map $n$ copies of it onto a single entangled two-qubit state by local operations and classical communication.} iff $\alpha\geq -1/2$ but unentangled only for $\alpha\geq -1/d$. Inside the gap that opens up for $d>2$, the boundary for $n$-copy undistillability is, so far, unknown even for $n=2$. Its notorious relevance stems from the fact that there exist so-called NPPT bound entangled states iff there is an $\alpha<-1/d$ for which $\rho_\alpha$ is   $n$-copy undistillable for all $n\in\N$. This problem has been shown to be equivalent to a set of $2$-norm inequalities for partial traces in \cite{Rico2025}. In particular, an entangled Werner state $\rho_\alpha$ is $2$-copy undistillable iff for every rank-$2$ matrix $M\in\cM_{d\otimes d}$:
\begin{equation}
    \norm{M_A}_2^2+\norm{M_B}_2^2\leq |\alpha|\big|\tr{M}\big|^2 +\frac{1}{|\alpha|}\norm{M}_2^2.\label{eq:2copyW}
\end{equation}
It was shown in \cite{Rico2025,PRAQi25} that Eq.(\ref{eq:2copyW}) holds (even without the trace term) for all $\alpha\in[-1/4,0)$. We can now extend this region as it follows from Cor.\ref{cor:AudenR} (when inserting $n=r=p=\gamma=2$) that 
\begin{equation}
    \norm{M_A}_2^2+\norm{M_B}_2^2\leq3\norm{M}_2^2.\label{eq:3}
\end{equation}
 Consequently:
\begin{corollary}
   Werner states are $2$-copy undistillable for all $\alpha\geq -1/3$. 
\end{corollary}
 Moreover, we know from the discussion of and after Cor.\ref{cor:AudenR} that the constant $3$ is best possible in Eq.(\ref{eq:3}). That is, any further extension of the interval of 2-copy undistillability must take the trace-term in Eq.(\ref{eq:2copyW}) into account.

\subsection{Application: Schmidt-number witnesses and $k$-positive maps}\label{sec:Schmidt}
We finally express some consequences of Cor.\ref{cor:AudenR} in the language of Schmidt-number witnesses and $k$-positive maps.

A Hermitian operator $W$ acting on a bipartite Hilbert space $\cH_A\otimes\cH_B$ is called a \emph{Schmidt-number witness} of class $(k+1)$ if $\langle\psi,W\psi\rangle\geq 0$ holds for all $\psi$ whose Schmidt-rank (i.e., the rank of the partial trace) is at most $k$ and $\langle\varphi, W\varphi\rangle<0$ for at least one $\varphi$ with Schmidt-rank $(k+1)$ \cite{Schmidtwitness}. This generalizes the concept of \emph{entanglement witnesses}, which are recovered as class $2$ Schmidt-number witnesses.

For $\cH_A=\bigotimes_{i=1}^n\cH_{A_i}\simeq\cH_B=\bigotimes_{i=1}^n\cH_{B_i}\simeq(\C^d)^{\otimes n}$ define $\omega_{(i)}:=\1\otimes\bigotimes_{j\in\{1,\ldots,n\}\setminus\{i\}}|\Omega_j\rangle\langle\Omega_j|$ on $\cH_A\otimes\cH_B$, where the identity acts on $\cH_{A_i}\otimes\cH_{B_i}$ and $|\Omega_j\rangle\in\cH_{A_j}\otimes\cH_{B_j}$ and 
\begin{equation}\label{eq:Wk}
    W_k:=\big(1+(n-1)k\big)\1-\sum_{i=1}^n \omega_{(i)}\;.
\end{equation}
\begin{corollary}
   For every $0\leq k<d$ Eq.(\ref{eq:Wk}) defines a Schmidt-number witness of class $(k+1)$.
\end{corollary}
\begin{proof}
    If we set $\psi=(M\otimes\1)|\Omega\rangle$  with $M$ acting on $\cH_A$ and $|\Omega\rangle=\bigotimes_{i=1}^n|\Omega_i\rangle$, then 
    \begin{equation}\label{eq:WK2}
        \langle\psi,W_k \psi\rangle=\big(1+(n-1)k\big)\norm{M}_2^2-\sum_{i=1}^n\norm{M_i}_2^2.
    \end{equation}
    Following Cor.\ref{cor:AudenR} this is non-negative for any $M$ of rank $k$. Moreover, we know from the discussion of sharpness of Cor.\ref{cor:AudenR} that Eq.(\ref{eq:WK2}) is negative for the rank $(k+1)$ matrix $M=|0\rangle\langle0|^{\otimes(n-1)}\otimes\1_{k+1}$.
\end{proof}
An alternative way to state this observation is in terms of $k$-\emph{positive maps}. A map $T$ between matrix spaces is called $k$-positive if $T\otimes\id_k$ is a positive map, where $\id_k:\cM_k\rightarrow\cM_k$ denotes the identity map. It is known \cite{THSchmidt} that a map $T$ is $k$-positive iff $\langle\psi,\big(T\otimes\id\big)\big(|\Omega\rangle\langle\Omega|\big)\psi\rangle\geq 0$ holds for all $\psi$ of Schmidt-rank $k$. This implies that we obtain a map that is $k$-positive when regarding $W_k$ from Eq.(\ref{eq:Wk}) as the Choi matrix of the map. This gives:
\begin{corollary}
    For $k,n\in\N$ the following map on $(\cM_d)^{\otimes n}$ is $k$-positive:
    \begin{equation}
        T(X):=\big(1+(n-1)k\big)\tr{X}\1-\sum_{i=1}^n \Tr_i[X]\otimes\1,
    \end{equation}
    where $Tr_i[X]\otimes\1$ denotes the partial trace of $X$ over the $i'$th tensor factor, with the traced-out factor being replaced by an identity.
\end{corollary}
\emph{Remark:} For $n=2$, a similar map $X\mapsto \tr{X}\1+X-\Tr_2[X]\otimes\1-\1\otimes \Tr_1[X]$ was shown to be $1$-positive by Ando \cite{ando2014matrix}.\vspace*{15pt}

\section*{Appendix: On computing $\kappa(c)$ for $p$-norms}\label{sec:appendix}
For the case of Schatten $p$-norms, and for the ease of notation $n=2$, following Thm.\ref{prop:Agen} and using that $\norm{\cdot}_p^*=\norm{\cdot}_q$ with $1/p+1/q=1$, we get
    \begin{equation}\label{eq:kappaschatten}
         \kappa(c)=\sup\bigg\{\sum_{i,j=1}^d(x_i+y_j-c)_+^q\;\Big|\; \sum_{i=1}^d |x_i|^q=\sum_{j=1}^d|y_j|^q=1,\; x,y\in\R^d\bigg\}^{1/q}.
    \end{equation}
The first observation is that  $\kappa(c)=0$ for $c\geq 2$. This follows from the constraint $\Vert x \Vert_q=\Vert y \Vert_q=1$ in \eqref{eq:kappaschatten} or from the fact that $\Vert M_i\Vert_p\leq \Vert M_i \Vert_1 \leq \Vert M\Vert_1$ for every $p\in [1,\infty]$. In addition, for $c\in [1,2]$,  it holds that $\kappa(c)=2-c$. Since $\kappa(c)$ is a convex function (as it is the supremum of convex functions),  $\kappa(1)=1$ by Prop.\ref{prop:ExtendedAudenaert} and $\kappa(2)=0$, it follows that $\kappa(c)\leq 2-c$. On the other hand, if we let $x=y=(1,0,\hdots,0)$, we obtain $\kappa(c)\geq 2-c$. For $c\in (0,1)$ the computation of the supremum becomes more intricate, but  we want to mention one observation that simplifies the computation or estimation of $\kappa(c)$  for other values of $n,p,d,r,c$ than discussed in the main text.

    At a maximum, which is clearly obtained for some $x,y\in\R_+^d$, there are Lagrange multipliers $\nu_1,\nu_2>0$ for which 
    \begin{eqnarray}\label{eq:x_ifunc}
        \left(\sum_{j=1}^d (x_i+y_j-c)_+^{q-1}\right)^{1/(q-1)} &=& \nu_1\; x_i\quad\forall i\\ \label{eq:y_ifunc}
        \left(\sum_{i=1}^d (x_i+y_j-c)_+^{q-1}\right)^{1/(q-1)} &=& \nu_2\; y_j\quad\forall j
    \end{eqnarray}
        Assume that $q> 2$. Seen as a function of a single $x_i$, the r.h.s. of Eq.(\ref{eq:x_ifunc}) is linear and increasing, while the l.h.s. is convex, constant zero for small values of $x_i$, possibly followed by an affine piece and, if $x_i$ is sufficiently large, a strictly convex part. Consequently, the graphs of the two functions can intersect in at most two points unless the affine part is linear. The latter, however, requires a fine-tuning of $y$ and is thus a case that can be ignored for the purpose of computing a supremum for reasons of continuity. We can therefore assume that the $x_i$'s and similarly all $y_j$'s take on at most two values.

        Hence, up to permuting coordinates, any candidate maximizer can be written 
\begin{equation}\label{eq:param}
x=(\underbrace{\alpha,\dots,\alpha}_{m\ \text{times}},\underbrace{\beta,\dots,\beta}_{d-m\ \text{times}}),
\qquad
y=(\underbrace{\gamma,\dots,\gamma}_{n\ \text{times}},\underbrace{\delta,\dots,\delta}_{d-n\ \text{times}}),
\end{equation}
with integers $1\le m,n\le d$ and parameters $\alpha>\beta\ge0$, $\gamma>\delta\ge0$ satisfying
\begin{equation}\label{eq:norm}
m\alpha^{q}+(d-m)\beta^{q}=1,
\qquad
n\gamma^{q}+(d-n)\delta^{q}=1.
\end{equation}

Similarly, for $1<q<2$ we can exploit Eqs.(\ref{eq:x_ifunc},\ref{eq:y_ifunc}) again to reduce the supports of candidate maximizers. However, as the l.h.s. of these equations is now a continuous function whose graph has a constant zero piece followed by a strictly concave piece, there are at most two non-zero intersections with a linear graph, but there might be a third one at zero.

While this does not completely solve the optimization problem for computing $\kappa(c)$, it significantly reduces the number of variables and, depending on the considered values of $n,p,d,r,c$, might lead to closed form solutions. \vspace*{15pt}

\emph{Acknowledgments:} The authors acknowledge funding by the Deutsche For-schungsgemeinschaft (DFG, German Research Foundation) under Germany's Excellence Strategy –  EXC-2111 – 390814868 and via the TRR 352 – Project-ID 470903074. MMW is grateful for the hospitality and support of the Cambridge Quantum Information group and of Pembroke College, Cambridge, where most of this work has been done.

\bibliographystyle{halpha}
\bibliography{PTI}{}\vspace*{15pt}

\end{document}